\documentclass{appolb}
\usepackage{epsfig}
\newcommand{\nuc}[2]{$^{#2}$#1}                 
\newcommand{\orb}[2]{${#1}_{{#2}/2}$}           
\begin{document}
\title{GLOBAL LIFETIME MEASUREMENTS OF HIGHLY-DEFORMED AND OTHER ROTATIONAL
STRUCTURES IN THE A$\sim$135 LIGHT RARE-EARTH REGION: PROBING THE
SINGLE-PARTICLE MOTION IN A ROTATING POTENTIAL%
\thanks{Invited talk presented at the ``High Spin Physics 2001" NATO
Advanced Research Workshop, dedicated to the memory of Zdzis\l{}aw Szyma\'nski,
Warsaw, Poland, February 6--10, 2001}
}%
\author{M.A. Riley,$^{1}$ R.W. Laird,$^{1}$  F.G. Kondev,$^{1,12}$
D.J. Hartley,$^{1,7}$ D.E. Archer,$^{1,13}$ T.B. Brown$,^{1}$ R.M.
Clark,$^{2}$ M. Devlin,$^{3}$ P. Fallon,$^{2}$  I.M.
Hibbert,$^{4}$ D.T. Joss,$^{5}$ D.R. LaFosse,$^{3}$ P.J.
Nolan,$^{5}$ N.J. O'Brien,$^{4}$ E.S. Paul,$^{5}$ J. Pfohl,$^{1}$
D.G. Sarantites,$^{3}$ R.K. Sheline,$^{1}$ S.L. Shepherd,$^{5}$
 J. Simpson,$^{6}$ R. Wadsworth,$^{4}$  M.T. Matev,$^{7}$
A.V. Afanasjev,$^{8,15}$  J. Dobaczewski,$^{9,10}$ G.A.
Lalazissis,$^{8,10,14}$ W. Nazarewicz,$^{7,9,11}$ and W.
Satu{\l}a$^{9,10}$
\address{
$^1$Department of Physics, Florida State University, Tallahassee, FL 32306
\\
$^2$Nuclear Science Division, Lawrence Berkeley National Lab., Berkeley, CA  94720
\\
$^3$Department of Chemistry, Washington University, St. Louis, MO 63130
\\
$^4$Department of Physics, University of York, Heslington, York Y01 5DD, UK
 \\
$^5$Oliver Lodge Laboratory, University of Liverpool, Liverpool L69 3BX, UK
\\
$^6$CLRC, Daresbury Laboratory, Daresbury, Warrington, WA4 4AD, UK
\\
$^7$Dept of Physics and Astronomy, Univ. of Tennessee, Knoxville, TN  37996
\\
$^8$Physik-Dept der Technischen Universit\"{a}t M\"{u}nchen \\ D-85747, Garching, Germany
\\
$^9$Institute of Theoretical Physics, Warsaw University \\ ul. Ho\.za 69,
PL-00681 Warsaw, Poland
\\
$^{10}$Joint Inststitute for Heavy Ion Research, Oak Ridge National Laboratory \\
P.O. Box 2008, Oak Ridge, TN 37831
\\
$^{11}$Physics Division, Oak Ridge National Laboratory, P.O. Box 2008, Oak
Ridge, TN 37831 \\
$^{12}$Physics Division, Argonne National Laboratory, Argonne, IL  60439 \\
$^{13}$Lawrence Livermore National Laboratory, Livermore, CA 94550 \\
$^{14}$Department of Theoretical Physics, Aristotle University of
Thessaloniki, Gr-54006 Thessaloniki, Greece \\
$^{15}$ Laboratory of Radiation Physics, Institute of Solid State
Physics, University of Latvia, LV 2169 Salaspils, Miera str. 31,
Latvia \\
}}
\maketitle
\vspace{-0.6cm}
\begin{abstract}
It has been possible, using {\small GAMMASPHERE} plus Microball,
to extract differential lifetime measurements free from common
systematic errors for over 15 different nuclei (various isotopes
of Ce, Pr, Nd, Pm, and Sm) at high spin within a single
experiment.  This comprehensive study establishes the effective
single-particle quadrupole moments in the A$\sim$135 light
rare-earth region. Detailed comparisons are made with calculations
using the self-consistent cranked mean-field theory.
\end{abstract}

\section{Introduction to Nuclear Shapes}

In 1936 Casimir argued that non-spherical nuclear shapes were
necessary to explain the hyperfine structure seen in atomic and
molecular spectra \cite{Cas36}.  A year later Bohr and Kalckar
\cite{BoKa} suggested that by observing the gamma-ray emissions
from excited nuclei one could find out more about these shape
properties.  Following these pioneering suggestions, the strongly
interacting aggregation of fermions in nuclei have been found to
display a remarkable diversity of both static and dynamic shape
phenomena (see for example References \cite{BoMo}, \cite{Szy83}
and \cite{RagNil}).

{\it Few scientists have contributed as much to the amazing
developments and excitement in our field as Szyma\'nski. I well
remember the first time I heard him speak. It was at one of the
famous May high spin gatherings at the Niels Bohr Institute in
Ris{\o} in the early 1980's. The discussion had become somewhat
bogged down on something rather trivial. It was then that
 Szyma\'nski rose to his feet from the back of the
audience, (in a similar manner to how his mind had just ``risen
above the tree-line in order to see the forest'') and speaking
softly in his wonderful way, he was able to steer the conversation
out of the ``trees" and towards something much more important. So
often we ``regular mortals" get carried away with the little
details and fail to comprehend the grand scheme or how it all fits
together. Szyma\'nski clearly possessed a rare ability to see the
``big picture". Amazingly, he has been able to pass this
leadership gift onto his ``scientific children''. His
achievements, and those of the Warsaw group which he created and
cultivated, are legendary. He was a true ``Titan'' of our field!}

The basic microscopic mechanism leading to the existence of
deformed configurations, {\em spontaneous symmetry-breaking}, was
first proposed by  Jahn and Teller for molecules \cite{Jah37}. The
basic element of the Jahn-Teller (JT) effect is the vibronic
coupling (the JT matrix element) between the collective
excitations of the system and the single-particle motion. The JT
effect was brought to nuclear structure by A. Bohr in his paper on
``The Coupling of Nuclear Surface Oscillations to the Motion of
Individual Nucleons"\,\cite{Boh52}. In Bohr's picture, the
vibronic coupling was represented by  the  interaction between the
single-particle motion of valence nucleons and the collective
excitations (multipole vibrations) of the core, known as
 the particle-vibration (PV) coupling.
This coupling
is a central element in the analysis of nuclear collective modes
and nuclear deformations \cite{BoMo}. Depending on the geometrical
properties of the individual valence nucleon (i.e., anisotropy of
its wave function), the PV coupling can result in core
polarization that can change the original deformation of the core.
If residual interactions  are present, they effectively reduce the
magnitude of the JT coupling. In particular, pairing correlations
in atomic nuclei give rise to a large residual energy (energy gap)
which weakens deformation effects. As a result, nuclear
ground-state configurations experience weak JT (or pseudo-JT)
effect; the extreme JT effect can take place in excited nuclear
states such as high spin states. (See Refs.~\cite{Rei84,Naz93} for
more discussion of the nuclear JT effect.)

\section{Nuclear Superdeformation, Rotation, and Quadrupole Moments}

A stunning observation in studies of
rapidly rotating atomic nuclei is that  due to the strong Coriolis
force residual   two-body correlations (e.g., pairing)
 are  significantly quenched at high angular momenta.
Indeed, the excellent reproduction of experimental high spin data
(moments of inertia, alignments, routhians) by
calculations without pairing (see, e.g., Refs. \cite{Afa99,Afa99a}
and references quoted therein) suggests that, in the
high-frequency regime, the configuration mixing due to pairing is
weak.  Another factor that further contributes to the diminished
role of residual correlations is
the reduced single-particle level density at the Fermi surface due to
the presence of strong shell effects at  deformed shapes. The
resulting deformed potentials  give rise to  magic numbers that
appear just as strikingly, but at quite different $N$
and $Z$ values, as for a spherical nucleus. Consequently,  highly
deformed and superdeformed bands  are probably the clearest
examples  of a pure single-particle motion in a deformed potential
\cite{deF96} and wonderful laboratories of the nuclear JT effect
and shape polarization.

In the $A$$\sim$135 ($Z$=58-62) light rare-earth region, a variety
of rotational sequences with characteristics consistent with
highly-deformed prolate shapes with quadrupole deformation of
$\beta_2$$\sim$0.30-0.40, compared with $\beta_2$$\sim$0.20 for
normal ground state deformations, have been observed
\cite{Firestone97}.  These studies have revealed that the
existence of highly-deformed bands is the result of an interplay
between microscopic shell effects, such as the occurrence of large
shell gaps near $Z$=58 and $N$=72 in the nucleon single-particle
energies, and the occupation of high-$j$ low-$\Omega$ intruder
orbitals. Initially, it was thought that the involvement of one or
more $i_{13/2}$ neutrons was necessary for the strong polarization
of the nuclear shape to stabilize large deformation in this mass
region \cite{Wy88}. However, recently it was shown that bands
involving the partial de-occupation of the extruder
$g_{9/2}$[404]9/2 proton orbital in the odd-$Z$ praseodymium (see
\cite{Ko99c} and references therein) and promethium \cite{pm133sd}
isotopes also exhibit comparable quadrupole
deformations\cite{Afan96}.
 Furthermore,
for nuclei below $N$=73 where the occupancy of the $i_{13/2}$
neutron is energetically unfavored, there are indications that
bands involving the $f_{7/2},h_{9/2}$ [541]1/2 neutron orbital may
also push the nucleus to ``enhanced'' deformation (see \cite{Ko99}
and references therein). (For representative single-particle
diagrams, see Refs.~\cite{Wy88,Afan96}.) In an attempt to further
understand the deformation properties of a variety of different
single-particle orbitals  throughout the $A$$\sim$135 mass region,
 a comprehensive lifetime experiment with measurements
on over 15 different nuclei was performed.  While such  an idea
 has been exploited for
several cases in the  $A$=80 \cite{Lerma99} and 135
\cite{Re92,Clark96} highly-deformed, and the $A$=150
\cite{Sav96,Ni97,Hackman98} and 190 \cite{Bus98} superdeformed
regions,  it has never been done before in such a global manner.

From the measured lifetimes, one can extract  ``differential''
transition quadrupole moments
\begin{equation}\label{deltaq}
\delta Q_{t} (^AZ;c) \equiv
 Q_{t} (^AZ;c) -  Q_{t}({\rm core}),
\end{equation}
where $c$ stands for the configuration of the  band in the nucleus
$^AZ$ and $Q_{t}({\rm core})$ is the transition quadrupole moment
of the  assumed core nucleus. According to
 the ``additivity principle"
proposed in Ref.~\cite{Sat96} (see also Ref.~\cite{Kar98}),
 the  quadrupole  moment $\delta Q_t$
can be expressed as a sum of individual contributions carried  by
individual particle and hole states which appear near the Fermi
level. Namely, the relative transition  quadrupole  moment can be
very well approximated by the ``extreme shell model" expression
\begin{equation}\label{dq}
\delta Q_{t} \approx \delta Q_{t}^{\rm SM} = \sum_{i}
 q_t(i),
\end{equation}
where $i$ runs over the particles and holes with respect to the
core  configuration in the nucleus $^AZ$. The quantity $q_t(i)$
represents the effective single-particle transition quadrupole
moment, i.e., the change of the total intrinsic moment which is
induced on the whole nucleus by the given particle or hole. By
measuring or calculating values of $Q_{t} (^AZ;c)$ for a number of
nuclei and configurations, one can extract values of $q_t(i)$
hence the quadrupole polarizabilities associated with  individual
orbitals.

The transition  quadrupole moments for various highly-deformed
structures in the $A$$\sim$135 region have been measured in
separate past experiments using the Doppler-shift attenuation
method (DSAM), however conclusive comparisons between similar
structures in different nuclei were limited because of systematic
differences such as reaction choice and target retardation
properties.  Specifically, the differences in the parameterization
of the nuclear and electronic stopping powers, which act as an
``internal clock'' in DSAM measurements, have contributed to large
variations in the extracted $Q_t$ values reported for exactly the
same nucleus and band. The absence of adequate experimental
information on the time structure of the quasi-continuum side
feeding contribution also provided additional uncertainty.  In the
present study, these systematic uncertainties were greatly reduced
because a variety of different nuclei were produced under similar
conditions and then analyzed using the same techniques.
Furthermore, the high efficiency and resolving power of {\small
GAMMASPHERE} coupled with Microball made it possible to explore
the time structure of the side-feeding into particular bands.

\section{Measurements}

In this study, high-spin states of a variety of $A$$\sim$135
nuclei ($Z$=58-62) were populated after fusion of a $^{35}$Cl beam
with an isotopically enriched 1 mg/cm$^{2}$ thick $^{105}$Pd foil
mounted on a 17 mg/cm$^{2}$ Au backing. The 173 MeV $^{35}$Cl beam
was provided by the 88-Inch Cyclotron at the Lawrence Berkeley
National Laboratory.  The emitted $\gamma$-rays were collected
using the {\small GAMMASPHERE} spectrometer~\cite{Ja96} consisting
of 97 Compton-suppressed HPGe detectors.  The evaporated charged
particles were identified with the Washington University Microball
detector system~\cite{Saran96} allowing a clean separation of the
different charged-particle exit channels.

\begin{figure}
\begin{center}
\includegraphics[scale=0.65]{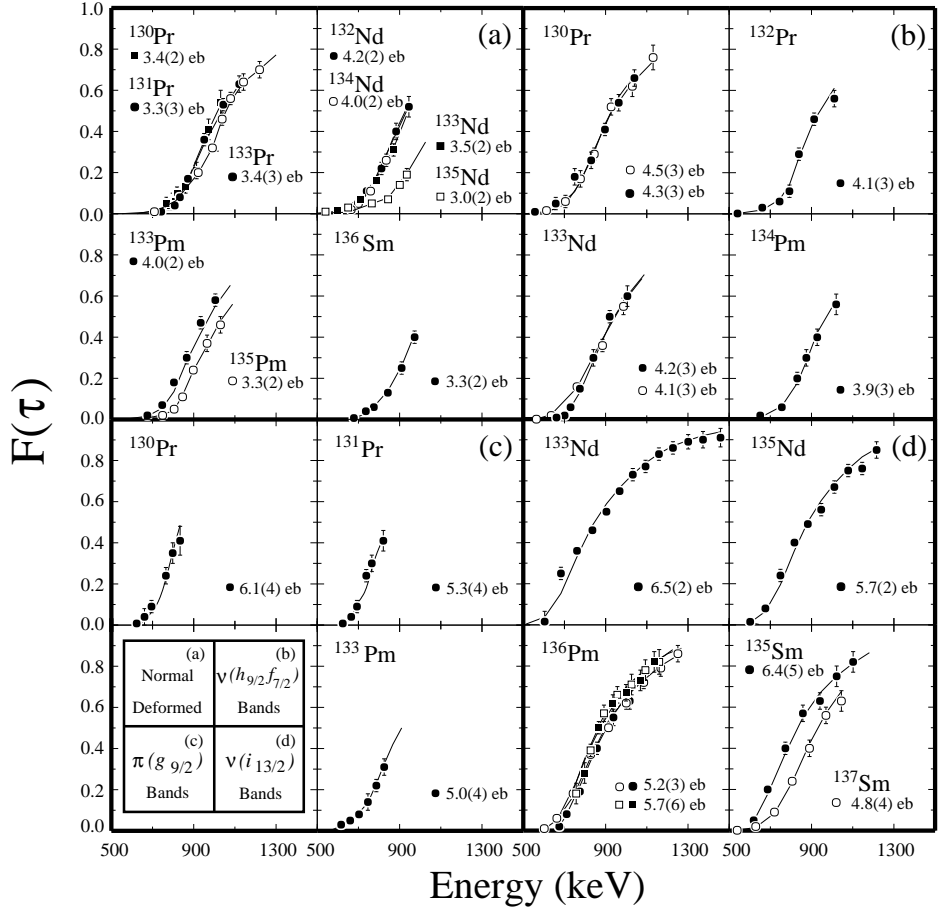}
\end{center}
\caption{
\noindent Global F($\tau$) curves for some of the (a)
normal deformed, (b) $\nu f_{7/2} h_{9/2}$, (c) $\pi g_{9/2}$, and
(d) $\nu i_{13/2}$ bands measured in this work. Extracted
quadrupole moment values are also given. } \label{ftau}
\end{figure}

The centroid-shift Doppler-shift attenuation method \cite{Alex78}
was used in several different ways. For the most intense bands,
spectra were generated by summing gates on the cleanest, fully
stopped transitions at the bottom of the band of interest and
projecting the events onto the ``forward'' (31.7$^{\circ}$ and
37.4$^{\circ}$) and ``backward'' (142.6$^{\circ}$ and
148.3$^{\circ}$) axes.  These spectra were then used to extract
the fraction of the full Doppler shift, $F(\tau)(=
\frac{E_{\gamma}(\theta) - E_{0}}{E_{0}\beta'cos(\theta)}$), for
transitions within the band of interest, [$E_{\gamma}(\theta)$ is
the centroid of the $\gamma$-ray energy distribution as measured
in a detector at angle $\theta$ with respect to the beam
direction, $E_{0}$ is the unshifted $\gamma$-ray energy, and
$\beta^{\prime}$ is the mid-target recoil velocity]. In addition,
for many bands double gates were set on in-band ``moving''
transitions in any ring of detectors and data were once again
incremented into separate spectra for events detected at
``forward'', ``90$^{\circ}$'', and ``backward'' angles.  For the
most intense bands, Doppler-shifted coincidence gates could also
be set on  the highest spin transitions within the band, making it
possible to eliminate the effect of side-feeding for states lower
in the cascade.  Approximately a 10\% increase in the deduced
quadrupole moment was found for the $i_{13/2}$ neutron
configurations when using the latter method compared to the value
extracted by gating on the stopped transitions at the bottom of
the same band, which is in agreement with recent $^{131,132}$Ce
results \cite{Clark96}. A sample of these results in presented in
Fig~\ref{ftau}.

\begin{figure}
\begin{center}
\includegraphics[scale=0.65]{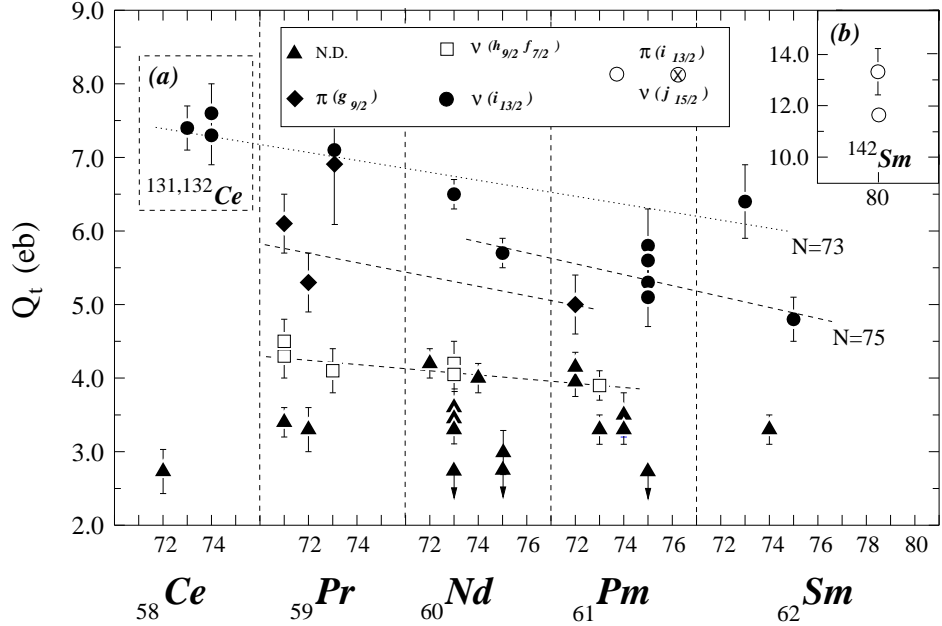}
\end{center}
\caption{
Plot, as a function of $Z$ and $N$,
summarizing the quadrupole moments, Q$_{t}$ extracted from this
experiment.  Quadrupole moments for the (a) highly-deformed bands
in $^{131,132}$Ce \protect \cite{Clark96} and (b) the
superdeformed bands in $^{142}$Sm \protect \cite{Hackman98} using
similar target and Au backing composition together with the same
analysis procedures and stopping powers as the current work.
}
\label{global}
\end{figure}

In order to extract the intrinsic quadrupole moments from the
experimental $F(\tau$) values, calculations using the code
{\scriptsize FITFTAU}~\cite{Moore96} were performed.  The
$F(\tau$) curves were generated under the assumption that the band
has a constant $Q_{t}$ value.
Although the uncertainties in the stopping powers and the modeling
of the side-feeding may contribute an additional systematic error
of 15$-$20$\%$ in the absolute $Q_{t}$ values, the relative values
are considered to be accurate to a level of 5$-$10$\%$.  Such
precision allows a clear differentiation between the quadrupole
polarizability of different orbital configurations for a variety
of $N$ and $Z$ values, see Fig~\ref{global} and Ref.\cite{Laird}.

\begin{table}[t]
\caption{ Transition quadrupole moments measured in this work.
Unless stated otherwise, $Q_t$ values were deduced assuming
$Q_{t}$(side feeding)=$Q_{t}$. The stopping power uncertainties
may contribute an additional systematic error of 10-15\% in the
absolute $Q_{t}$ values. }
\label{tableex}
\begin{center}
\begin{tabular}{ccc}
Nucleus & Configuration$^a$ & $Q_t$(eb) \\
\hline
$^{130}$Pr & $\pi$\protect\orb{h}{11}$\otimes$$\nu$\protect\orb{d}{5} & 3.4(2) \\
   & $\pi$\orb{h}{11}$\otimes$$\nu$(\orb{f}{7},\orb{h}{9})~(band 1)/(band 2) & 4.3(3)/4.5(3)\\
      & $\pi$\orb{g}{9}$\otimes$$\nu$\orb{h}{11} & 6.1(4)\\
\protect\nuc{Pr}{131} & $\pi$\orb{h}{11} & 3.3(3) \\
                      & $\pi$\orb{g}{9}  & 5.3(4) \\
\nuc{Pr}{132} & $\pi$\orb{h}{11}$\otimes$$\nu$(\orb{f}{7},\orb{h}{9}) & 4.1(3)\\
& $\pi$\orb{g}{9}$\otimes$$\nu$\orb{i}{13} & 7.0(7)\\
\nuc{Pr}{133} & $\pi$\orb{h}{11}
 & 3.3(3) \\
 \nuc{Nd}{132} & $\pi$\orb{h}{11}
 & 4.2(2) \\
 \nuc{Nd}{133}& $\nu$\orb{h}{11}
  & $<$ 3.0 \\
              & $\nu$\orb{h}{11}$ \otimes$ $\pi$(\orb{h}{11})$^{2}$
& 3.5(3)\\
             & $\nu$\orb{g}{7}  ($\alpha$ = +1/2)/($\alpha$ = --1/2) & 3.4(2)/3.5(2) \\
& $\nu$(\orb{f}{7},\orb{h}{9}) ($\alpha$ = +1/2)/($\alpha$ =--1/2) & 4.2(3)/4.1(3) \\
              & $\nu$\orb{i}{13}
& 5.8(2),6.5(2)$^{b}$ \\
\nuc{Nd}{134} & $\pi$\orb{h}{11}
& 4.0(2) \\
 \nuc{Nd}{135}& $\nu$\orb{h}{11}
& $<$ 3.0  \\
   & $\nu$\orb{h}{11}$ \otimes$ $\pi$(\orb{h}{11})$^{2}$
& 3.0(3) \\
& $\nu$\orb{i}{13}& 5.1(2),5.7(2)$^{b}$ \\
\hline
\end{tabular}
\end{center}

$^{a}$$\pi$\orb{g}{9}: 9/2$^{+}$[404],
$\pi$\orb{h}{11}: 3/2$^{-}$[541], $\pi$\orb{d}{5}: 3/2$^{+}$[411],
$\nu$\orb{h}{11}: 7/2$^{-}$[523] and 9/2$^{-}$[514],
$\nu$\orb{g}{7}: 7/2$^{+}$[404], $\nu$\orb{d}{5}: 5/2$^{+}$[402],
$\nu$(\orb{f}{7},\orb{h}{9}): 1/2$^{-}$[541], $\nu$\orb{i}{13}:
1/2$^{+}$[660].

$^{b}$Deduced by gating above the level of interest, so that
side-feeding was eliminated. One may reasonably expect a similar
increase of $\sim$10\% for the other highly-deformed bands listed
\cite{Ko99b}.
\end{table}

\setcounter{table}{0}
\begin{table}[t]
\caption{Continued.}
\begin{center}
\begin{tabular}{ccc}
Nucleus & Configuration$^a$ & $Q_t$(eb) \\
\hline
\nuc{Pm}{133} & $\pi$\orb{h}{11}
 & 4.0(2) \\
 & $\pi$\orb{d}{5}
  & 4.1(2) \\
               & $\pi$\orb{g}{9}
  & 5.0(4) \\
\nuc{Pm}{134} & $\pi$\orb{h}{11}$\otimes$$\nu$\orb{h}{11}
& 3.3(2)\\
  & $\pi$\orb{h}{11}$\otimes$$\nu$(\orb{f}{7},\orb{h}{9})
& 3.9(2)\\
\nuc{Pm}{135} & $\pi$\orb{h}{11}& 3.3(2)  \\
               & $\pi$\orb{d}{5}
& 3.5(2)  \\
\nuc{Pm}{136} & $\pi$\orb{h}{11}$\otimes$$\nu$\orb{h}{11}
 & $<$ 3.0 \\
              &   $\pi$\orb{h}{11}$\otimes$$\nu$\orb{i}{13} (band 1)
& 4.8(3),5.2(3)$^{b}$  \\
 &   $\pi$\orb{h}{11}$\otimes$$\nu$\orb{i}{13} (band 2)
& 4.8(4),5.2(4)$^{b}$ \\
  &   $\pi$\orb{g}{7}$\otimes$$\nu$\orb{i}{13} (band 1)
& 5.7(6) \\
 &   $\pi$\orb{g}{7}$\otimes$$\nu$\orb{i}{13} (band 2)
& 5.7(6) \\
\nuc{Sm}{135} &   $\nu$\orb{i}{13}
 & 5.8(4),6.4(4)$^{c}$ \\
\nuc{Sm}{136} &  $\nu$(\orb{h}{11})$^{2}$ $\otimes$ $\pi$(\orb{h}{11})$^{2}$ & 3.3(2)\\
\nuc{Sm}{137} &   $\nu$\orb{i}{13}
 & 4.4(3),4.8(4)$^{b}$ \\
\hline
\end{tabular}
\end{center}

$^{a}$$\pi$\orb{g}{9}: 9/2$^{+}$[404],
$\pi$\orb{h}{11}: 3/2$^{-}$[541], $\pi$\orb{d}{5}: 3/2$^{+}$[411],
$\nu$\orb{h}{11}: 7/2$^{-}$[523] and 9/2$^{-}$[514],
$\nu$\orb{g}{7}: 7/2$^{+}$[404], $\nu$\orb{d}{5}: 5/2$^{+}$[402],
$\nu$(\orb{f}{7},\orb{h}{9}): 1/2$^{-}$[541], $\nu$\orb{i}{13}:
1/2$^{+}$[660].

$^{b}$Deduced by gating above the level of interest, so that
side-feeding was eliminated. One may reasonably expect a similar
increase of $\sim$10\% for the other highly-deformed bands listed
\cite{Ko99b}.

$^{c}$Assumed that the sidefeeding time
 structure was similar to other highly-deformed $\nu$\orb{i}{13}
 bands where the $\tau$$_{sf}$$\sim$1.2$\tau$$_{band}$.
\end{table}

The present results, displayed in Table~\ref{tableex}, especially
when taken together with values for \nuc{Ce}{131,132}
\cite{Clark96}, extracted using an identical analysis procedure,
clearly indicate that for structures involving the
$\nu$\orb{i}{13} orbital there is a systematic decrease in the
deformation as a function of increasing proton and neutron
numbers, see Fig~\ref{global}.

Another important finding involves the highly-deformed bands built
upon the 9/2$^{+}$[404] (\orb{g}{9}) proton orbital in
\nuc{Pr}{130,131} and \nuc{Pm}{133} \cite{Ga94,pm133sd,Br97}. Our
results confirm the important role played by the
$\pi$\orb{g}{9}[404]9/2$^{+}$ hole orbitals in building
highly-deformed structures in this region. It is seen in
Fig~\ref{global} and Table~\ref{tableex} that the largest
deformations are observed for the Ce isotopes ($Z$$=$58) where the
9/2$^{+}$[404] orbital lies above the Fermi surface \cite{Afan96}.

It is also of interest to see by how much the strongly downsloping
1/2$^{-}$[541] (\orb{f}{7},\orb{h}{9}) neutron orbital polarizes
the nucleus towards an ``enhanced'' deformation. The trends for
this orbital, which has $Q_{t}$ values intermediate between the
highly-deformed ($\nu$\orb{i}{13} and $\pi$\orb{g}{9}) and normal
deformed configurations, also seems to exhibit slightly decreasing
$Q_{t}$ values as a function of increasing $Z$. The normal
deformed structures which dominate the yrast lines of these nuclei
for low spin values, display $Q_{t}$ values distinctly lower that
the above mentioned configurations, with a generally constant
trend as a function of $Z$ and $N$.  However, since these normal
deformed structures are predicted to be extremely $\gamma$-soft
and strongly influenced by pairing correlations,
 one has to exercise caution
when interpreting trends  in terms of single-particle occupations.

\section{Theoretical Calculations and Differential Q$_t$ Values}

To obtain quantitative understanding of measured quadrupole
moments, we performed systematic  cranking calculations without
pairing using two different self-consistent mean-field methods,
namely the cranked Skyrme Hartree-Fock method (CSHF) (code {\sc
HFODD} \cite{hfodd}) with the Skyrme parameterization  SLy4
\cite{Cha98}, and the cranked relativistic mean-field theory
(CRMF) \cite{KR89,AKR96} with the  parameterization NL1
\cite{NL1}. Both methods have shown to  provide an accurate
description of various properties of rotational bands in different
mass regions (see, e.g., Refs. \cite{Sat96,Aou00,AKR96,Afa99}).
For the details pertaining to theoretical calculations, see the
forthcoming Ref.~\cite{Matev}.

For the reference band, we took the lowest highly deformed $\nu
(i_{13/2})$  intruder band in $^{131}$Ce. According to
calculations \cite{Wy88},
 large deformed energy gaps develop at high angular
momentum  for $Z$=58 and  $N$=73, i.e.,  $^{131}$Ce can be
considered as a (super)deformed core in the $A$$\sim$135 mass
region. In order to perform a reliable statistical analysis of
individual quadrupole moments according to Ref.~\cite{Sat96}, it
was necessary to carry out  calculations for a  large number of
nuclei and configurations: our data set consisted of over 100
bands in both CSHF and CRMF calculations.

\begin{table}[t]
\caption{Effective charge quadrupole moments $q_{20}$ (in eb) for
single-particle orbitals around \nuc{Ce}{131} core. The values
were extracted from the set of 183 calculated bands in CSHF and
105 bands in CRMF. The orbitals are labeled by means of asymptotic
quantum numbers $[Nn_z\Lambda]\Omega$ and the signature quantum
number $\alpha$ (the subscripts $\pm$ stand for
$\alpha$=$\pm$1/2). Note that in most cases the signature
dependence is very weak (as observed in experiment, see Table 1).
} \label{qmom}
\begin{center}
\begin{tabular}[t]{ccc}
Orbital  & CSHF & CRMF \\
\hline
     \multicolumn{3}{c} {Neutrons}\\
$[402]\frac{5}{2}_{-}$  &      --0.35 &      --0.26\\
$[402]\frac{5}{2}_{+}$  &      --0.33 &      --0.26\\
 $[411]\frac{1}{2}_{-}$  &      --0.15 &      --0.11\\
 $[411]\frac{1}{2}_{+}$  &      --0.12 &      --0.06\\
 $[411]\frac{3}{2}_{-}$  &      --0.15 &      --0.13\\
 $[411]\frac{3}{2}_{+}$  &      --0.11 &      --0.12\\
 $[413]\frac{5}{2}_{-}$  &      --0.13 &      --0.13\\
 $[413]\frac{5}{2}_{+}$  &      --0.12 &      --0.11\\
 $[523]\frac{7}{2}_{-}$  &       0.03 &       0.05\\
 $[523]\frac{7}{2}_{+}$  &       0.04 &       0.01\\
 $[530]\frac{1}{2}_{-}$  &       0.22 &       0.17\\
 $[530]\frac{1}{2}_{+}$  &       0.17 &       0.19\\
 $[532]\frac{5}{2}_{-}$  &       0.19 &       0.17\\
 $[532]\frac{5}{2}_{+}$  &       0.24 &       0.38\\
 $[541]\frac{1}{2}_{-}$  &       0.35 &       0.35\\
 $[541]\frac{1}{2}_{+}$  &       0.37 &       0.33\\
 $[660]\frac{1}{2}_{+}$  &       0.38 &       0.40\\
 $[660]\frac{1}{2}_{-}$  &       0.36 &       0.36\\
\end{tabular}
~~~~\begin{tabular}[t]{ccc}
Orbital  & CSHF & CRMF \\
\hline
     \multicolumn{3}{c} {Protons}\\
$[404]\frac{9}{2}_{-}$  &      --0.32 &      --0.37\\
$[404]\frac{9}{2}_{+}$  &      --0.32 &      --0.37\\
$[422]\frac{3}{2}_{-}$  &       0.33 &       0.33\\
$[422]\frac{3}{2}_{+}$  &       0.34 &       0.28\\
$[532]\frac{5}{2}_{-}$  &       0.43 &       0.41\\
$[532]\frac{5}{2}_{+}$  &       0.56 &       0.54\\
$[541]\frac{3}{2}_{-}$  &       0.50 &       0.48\\
$[541]\frac{3}{2}_{+}$  &       0.57 &       0.50\\
$[550]\frac{1}{2}_{+}$  &       0.49 &       0.47\\
\end{tabular}
\end{center}
\end{table}

\begin{figure}
\begin{center}
\includegraphics[scale=0.50]{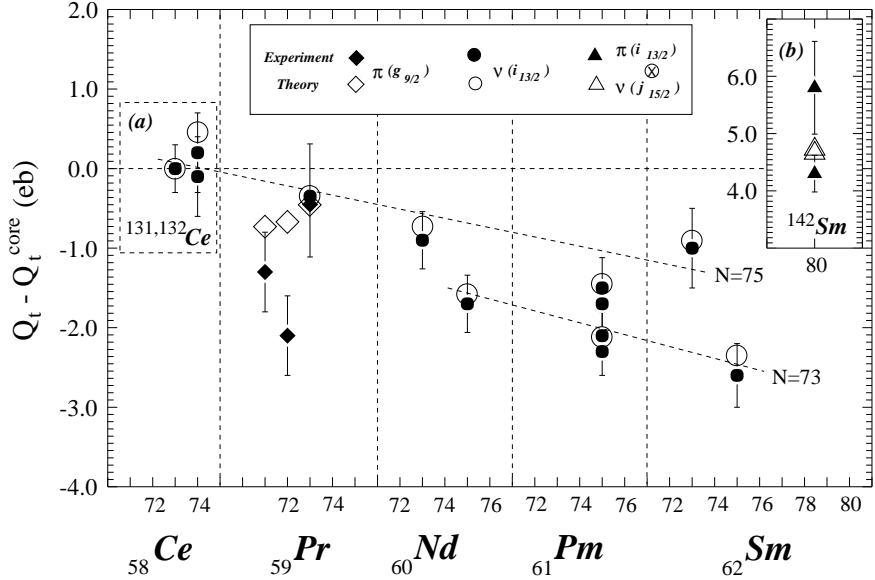}
\end{center}
\caption{
\noindent Experimental (closed symbols with error bars)
 and calculated (CSHF; open symbols) differential transition quadrupole moments
 (equation \protect\ref{deltaq}) for a number of nuclei from the $A$$\sim$135
mass region. The highly-deformed band in $^{131}$Ce (containing
one \orb{i}{13} neutron and two \orb{g}{9} proton holes) was used
as a reference core configuration. The data for  \nuc{Ce}{131,132}
were taken from Ref.~\protect\cite{Clark96}. The corresponding
values for $^{142}$Sm (see Ref.~\protect\cite{Hackman98} for
experimental data) are shown in the inset. It is to be noted that
for the differential quadrupole moments the problem with the
stopping-power uncertainties is less severe. For comparison with
Fig. 2, the calculated value of  $Q_t$ for the $^{131}$Ce core
configuration was  7.75~eb in CSHF and 7.57~eb in CRMF. }
\label{Qtfig}
\end{figure}

Within the cranking model, the transition quadrupole moment can be
written as \cite{Ring82}
\begin{equation}\label{Qth}
Q_t =Q_{20} + \sqrt{2\over 3} Q_{22},
\end{equation}
where $Q_{20}$ and $Q_{22}$ are calculated components of the
quadrupole moment. (Let us note that due to the non-standard
normalisation of $Q_{22}$ in Ref.~\cite{hfodd}, one gets
$Q_{22}$(Ref.~\cite{hfodd})=$-\sqrt{2}Q_{22}$.) Figure~\ref{Qtfig}
shows the experimental and calculated (CSHF) values of $Q_t$
relative to a $^{131}$Ce core, for the highly deformed bands in
nuclei with $Z$=57-62 involving \orb{i}{13} neutrons  and/or
\orb{g}{9} proton holes. The agreement  between experiment and
theory is quite remarkable, with all the major experimental trends
discussed previously being well reproduced.

{}From the differential proton quadruple moments $\delta
Q^{\pi}_{20}$ and $\delta Q^{\pi}_{22}$ calculated at  rotational
frequency of
 $\hbar\omega$=0.65\,MeV, effective single-particle  charge quadrupole moments
$q_{20}(i)$ and $q_{22}(i)$ were extracted for protons and
neutrons according to the additivity principle, Eq.~(\ref{dq}). Of
course, $q_t(i)=q_{20}(i)+\sqrt{2\over 3} q_{22}(i)$ are
 the effective values for the single-particle/hole states.
The results of our linear regression analysis for $q_{20}$ are
displayed in Table~\ref{qmom}. (For tabulated values of
$q_{22}$ and angular momentum alignments, see Refs.~\cite{Matev}.)
It is seen that the two models give very similar results. In
particular, quadrupole polarizabilities of the lowest \orb{i}{13}
neutron orbitals, [541]1/2 neutron orbitals,
 and \orb{g}{9} proton holes
are comparable. The fact that the  bands attributed to the
[541]1/2 neutron orbital are experimentally significantly less
deformed than the \orb{i}{13}  intruder structures can probably be
attributed to pairing correlations which effectively reduce the
occupation of [541]1/2. With these values at hand one can
calculate quadrupole moments $Q_{20}$ for near-axial bands in the
$A$$\sim$135 mass region.

The general trend of decreasing $Q_t$ in the highly-deformed
structures with increasing $Z$ and $N$, see Fig.~\ref{Qtfig},  is
consistent with general expectations that as one adds particles
above a deformed shell gap, the stabilizing effect of the gap may
be diminished. This trend continues until a new ``magic'' deformed
number is reached.  Such an event clearly occurs from
\nuc{Ce}{132} to higher $N$ and $Z$ until $Z$=62 and $N$=80
(\nuc{Sm}{142}) where a large jump in quadrupole moment occurs
marking the point at which it becomes energetically favorable to
fill the high-$j$ $\pi$\orb{i}{13} and $\nu$\orb{j}{15} orbitals
creating the A$\sim$142 superdeformed island, see the inset in
Fig.~\ref{Qtfig}. It is gratifying to see that theory can
reproduce the $Q_t$ in \nuc{Sm}{142} using {\em both}
\nuc{Ce}{131} and  \nuc{Dy}{152} cores.

\section{Summary and Conclusions}

In summary, it has been possible to extract differential
transition quadrupole moments,
 free from common systematic errors
for over 15 different nuclei at high spin within a single
experiment.  This comprehensive study establishes $Z$, $N$, and
configuration dependent quadrupole moment trends in the A$\sim$135
light rare-earth region.  Detailed comparisons are made with
theoretical calculations using the Cranked Hartree-Fock and
Cranked Relativistic Mean Field frameworks. Theoretical
differential  transition quadrupole moments agree very well with
experimental data for highly deformed intruder bands in this
region. Based on the additivity principle, valid in the limit of
weak residual correlations, the effective single-particle charge
quadrupole moments  have been  obtained. Together with values of
Ref.~\cite{Sat96} around  \nuc{Dy}{152}, they can be used to
estimate the quadrupole moments of specific near-axial
configurations in a wide range of nuclei.

\bigskip
Discussions with R.V.F. Janssens, D. Ward, and A. Galindo-Uribarri
are acknowledged and greatly appreciated. Special thanks to D.C.
Radford and H.Q. Jin for software support, and to R. Darlington
for help with the target. The authors wish to extend their thanks
to the staff of the LBNL {\small GAMMASPHERE} facility for their
assistance during the experiment.  This work was supported in part
by the U.S. Department of Energy under Contract Nos.\
DE-FG02-96ER40963 (University of Tennessee), DE-FG05-87ER40361
(Joint Institute for Heavy Ion Research), and DE-AC05-00OR22725
with UT-Battelle, LLC (Oak Ridge National Laboratory), the
National Science Foundation, the State of Florida, the U.K.
Engineering and Physical Science Research Council, the U.K.
Council for the Central Laboratory of the Research Councils, and
the Polish Committee for Scientific Research (KBN).  MAR and JS
acknowledge the receipt of a NATO Collaborative Research Grant.
A.V.A. acknowledges support from the Alexander von Humboldt
Foundation.

%
%

\end{document}